\documentclass{ifacconf}

\usepackage{graphicx}      
\graphicspath{{figures/}}

\usepackage{amsfonts}
\usepackage{amsmath}
\usepackage[sort&compress,square]{natbib}        
\begin{document}
\begin{frontmatter}

\title{Constructing a control-ready model of EEG signal during general anesthesia in humans\thanksref{footnoteinfo}} 

\thanks[footnoteinfo]{This research was supported by the National Institutes of Health grants T32 HLO9701 and F32 AG064886 (to JHA). The authors thank Gabriel Schamberg for helpful comments. This work has been submitted to IFAC for possible publication.}

\author[1,2,3]{John H. Abel} 
\author[1,2]{Marcus A. Badgeley} 
\author[4]{Taylor E. Baum} 
\author[1,2]{Sourish Chakravarty} 
\author[1,2]{Patrick L. Purdon} 
\author[1,2,4,5,6]{Emery N. Brown} 

\address[1]{Department of Anesthesia, Critical Care and Pain Medicine, Massachusetts General Hospital, Boston, MA 02114}
\address[2]{Picower Institute for Learning and Memory, Massachusetts Institute of Technology, Cambridge, MA 02139}
\address[3]{Division of Sleep Medicine, Harvard Medical School, Boston, MA 02115}
\address[4]{Department of Brain and Cognitive Sciences, Massachusetts Institute of Technology, Cambridge, MA 02139}
\address[5]{Institute of Medical Engineering and Science, Massachusetts Institute of Technology, Cambridge, MA 02139}
\address[6]{enb@neurostat.mit.edu}

\begin{abstract}                
Significant effort toward the automation of general anesthesia has been made in the past decade. One open challenge is in the development of control-ready patient models for closed-loop anesthesia delivery. Standard depth-of-anesthesia tracking does not readily capture inter-individual differences in response to anesthetics, especially those due to age, and does not aim to predict a relationship between a control input (infused anesthetic dose) and system state (commonly, a function of electroencephalography (EEG) signal). In this work, we developed a control-ready patient model for closed-loop propofol-induced anesthesia using data recorded during a clinical study of EEG during general anesthesia in ten healthy volunteers. We used principal component analysis to identify the low-dimensional state-space in which EEG signal evolves during anesthesia delivery. We parameterized the response of the EEG signal to changes in propofol target-site concentration using logistic models. We note that inter-individual differences in anesthetic sensitivity may be captured by varying a constant cofactor of the predicted effect-site concentration. We linked the EEG dose-response with the control input using a pharmacokinetic model. Finally, we present a simple nonlinear model predictive control \textit{in silico} demonstration of how such a closed-loop system would work.
\end{abstract}


\end{frontmatter}

\section{Introduction}
\begin{figure}
\begin{center}
\includegraphics[width=8.4cm]{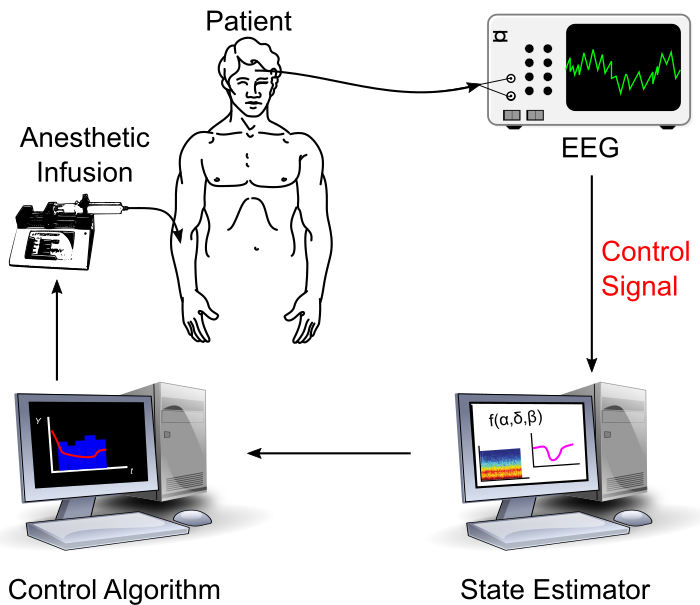}    
\caption{Diagram of a closed-loop anesthesia delivery (CLAD) system. This work focuses on development of the control signal appropriate for closed-loop delivery of propofol.} 
\label{fig:clad}
\end{center}
\end{figure}

General anesthesia (GA) is a profound state of unconsciousness, analgesia, and amnesia \citep{Brown2010}.  
Anesthesiologists control the depth of unconsciousness during GA by assessing patient signs and symptoms and/or using a brain monitor summary score and then making concomitant changes to the infusion of hypnotic medications at irregular time intervals.
Automatic control of unconsciousness with closed loop anesthesia delivery (CLAD) would afford anesthesiologists more attention on other aspects of anesthesia care and provide more regular and frequent adjustments to hypnotic infusions.
A diagram of a CLAD system is shown in Figure~\ref{fig:clad}.

Clinically available depth of unconsciousness summary monitors have been used as the control signal for prior studies on CLAD \citep{Absalom2009, Soltesz2012, VanHeusden2014, Gentilini2001, Haddad2003} as well as commercially available CLAD products \citep{Liu2015}.
Commercially available consciousness summary monitors include bispectral index (BIS, Medtronic) and NeuroSENSE (Neurowave Systems).
These monitors indicate the ``depth of anesthesia'' with a single scalar between 0 and 100.
Patients have been shown to have inter-individual differences in both pharmcodynamic (sensitivity) and pharmacokinetic (drug uptake and elimination) monitor responses to hypnotic \citep{Gentilini2001, Absalom2009}, particularly for children \citep{Soltesz2012, VanHeusden2014}.
Depth of anesthesia summary monitors are insensitive to patient characteristics and anesthetic agents, and defining the underlying physics of general anesthesia remains a significant barrier for CLAD \citep{Absalom2011a}.

The summary scores produced by consciousness monitors are dervied from recordings of the brain's electrical activity: the electroencephalogram (EEG).
Custom metrics dervied from the EEG have been used for automatic control of medically-induced coma.
Specifically, the probability of brain activity senescence (the "burst suppression probability") was computed from EEG in real-time and used as the control signal for developing linear-quadratic regulator \citep{Shanechi2013, Yang2019} and proportional-integral-derivative \citep{Ching2013} medically-induced coma controllers.
Medically-induced coma is only indicated for patients with aberrant brain activity, and targets an even greater reduction in brain activity than GA.
EEG can provide a richer representation of neural activity than summary monitors, but an EEG signature would need to be identified specifically for GA.

CLAD models have a wide-range of personalization and adaptability. 
Many fit model parameters on patient populations \citep{Absalom2009, Soltesz2012, VanHeusden2014, Gentilini2001}.
Some have developed personalization via an initial calibration period prior to control tests \citep{Shanechi2013, Ching2013}.
Calibration during control has been previously performed for medically-induced coma \citep{Yang2019} and in numerical simulations of general anesthesia \citep{Haddad2003}.

Here, we develop a proof-of-concept EEG-based CLAD system for general anesthesia.
We first derive an EEG-based control signal from a clinical trial on EEG response to the hypnotic agent propofol.
We then assess PK and PD model fitting differences across patients.
Next we formalize the application of a model-based controller with a adaptive phase of control that adjusts to an individual’s susceptibility.
Finally, we perform simulations to demonstrate the proposed system’s performance.  

\section{Problem Formulation and Approach}

We seek to ensure unconsciousness during anesthesia by regulating neural activity as recorded via EEG.
To do so, we aimed to develop a pharmacokinetic-pharmacodynamic (PKPD) model of EEG power spectral density (PSD) that evolves over time $t$ according to:
\begin{equation}
\frac{dp}{dt} = g(t, p, u)
\end{equation}
where system $p\in\mathbb{R}^D$ is some $D$-dimensional measure of the relevant EEG characteristics and control input $u\in\mathbb{R}_{\geq 0}$ is the anesthetic infusion dosing (concentration/time).
By noting that the PSD is a function of effect-site concentration of the drug
\begin{equation}
p \equiv p(c(t))
\end{equation} 
the model may be separated into two components:
\begin{equation}
\frac{dp}{dt} = \left(\frac{dp}{dc}\right)\left(\frac{dc}{dt}\right)
\end{equation}
where $c(t)$ is the drug concentration at the effect site.
The latter term, $dc/dt$, is the pharmacokinetic term and has been studied extensively and implemented clinically in the target-controlled infusion (TCI) paradigm \citep{Schnider1998, Schnider1999, Levitt2005, Barakat2007, Absalom2009}.
In this work, we focus on developing the former pharmacodynamic term, $dp/dc$, in a manner that enables control and is robust to inter-individual variability in drug responses.

\subsection{Identification of low-dimensional state space model from EEG}

We developed a control signal using the EEG data recorded from prefrontal cortex (Fp1 electrode) during a clinical trial of ten healthy volunteers where the effect-site concentration of propofol was varied systematically for each individual.
For details on the study and data collection, see \cite{Purdon2013}.
Because the power spectral density is known to vary reliably depending upon propofol-induced anesthetic state, we first performed a multitapered spectral analysis of the full Fp1 EEG time series to arrive at the multitaper spectrogram $M_s\in\mathbb{R}^{F\times N}$, where $F$ is the number of frequency bins and $N$ is the number of time windows of the spectrogram.
Here, the subscript $s$ denotes the individual subject from which the data was collected.
Each column of $M_s$ is denoted $m_{s,t}\in\mathbb{R}^{F\times 1}$ is the PSD in each of $F$ frequencies at time window $t$ for subject $s$.

We performed multitaper spectral analysis in Python using the NiTime package \citep{nitime} using window length $2$s with no overlap, normalized half-bandwidth $TW=3$, and a spectral resolution of $2$ Hz. An adaptive weighting routine was used to combine estimates of different tapers \citep{Thomson1982}, and the resulting multitaper spectrogram from each patient was converted to decibels.

\begin{figure}
\begin{center}
\includegraphics[width=8.4cm]{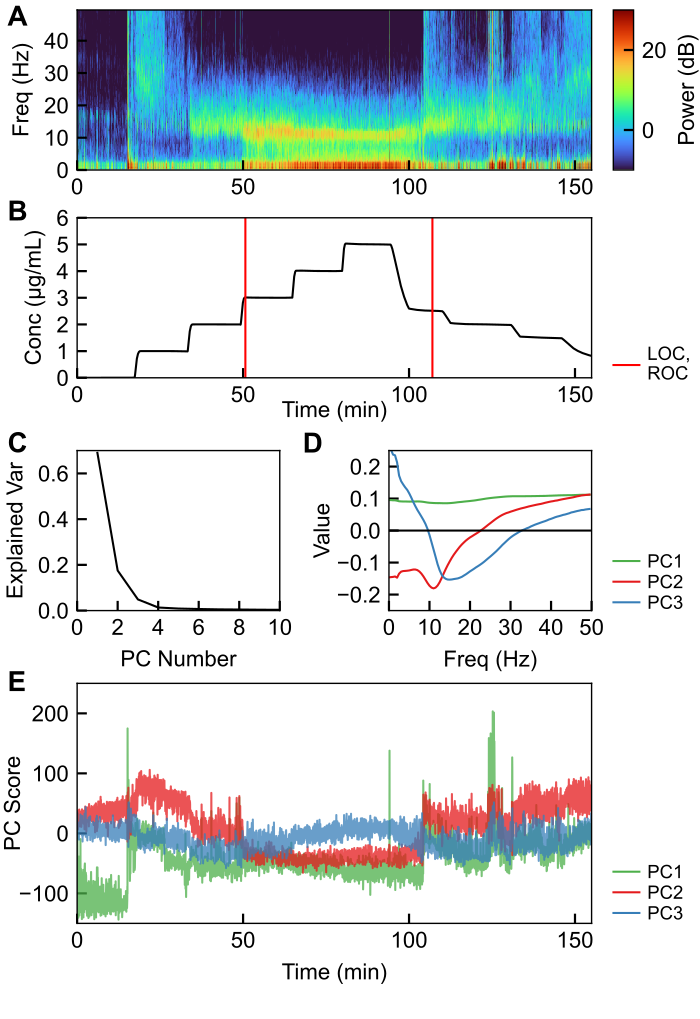}    
\caption{Spectral characteristics of EEG during propofol anesthesia. (A) Multitaper spectrogram of EEG signal in a healthy volunteer, from \cite{Purdon2013}. (B) Drug effect-site concentration corresponding to the multitaper spectrogram in A, with loss of consciousness (LOC) and restoration of consciousness (ROC) labeled. (C) Fraction of the explained variance in each of the first ten principal components. (D) Eigenvectors $e_1$, $e_2$, and $e_3$ corresponding to the first three principal components. (E) First three principal component score of the multitaper spectrogram furing the timecourse in (A,B).} 
\label{fig:pca}
\end{center}
\end{figure}

The manifold along which the EEG PSD evolves during anesthesia is of reduced dimensionality compared to $F$, and many spectral features co-occur, e.g., slow-delta (0.1-4 Hz) and alpha (8-12 Hz) \citep{Purdon2013}.
Thus, we sought to reduce the dimensionality of the observations to independent linear combinations of spectral power via projecting the PSD (each $m_{s,t}$) onto $D$ principal components.
That is, we compressed $M_s\in\mathbb{R}^{F\times N}$ to $P_s\in\mathbb{R}^{D\times N}$ by:
\begin{equation}
p_{i,s,t} = e_i \cdot m_{s,t}
\end{equation}
where $e_i \in\mathbb{R}^{1\times F}$ is the $i$th principal component, ordered by the magnitude of the corresponding eigenvalue.
We computed the principal components performing PCA on all ten resulting spectrograms $[M_1, \cdots, M_S]$, thus, the principal components are not a signature specific to each subject.
The result of this approach is shown in Figure~\ref{fig:pca}.

\begin{figure}
\begin{center}
\includegraphics[width=8.4cm]{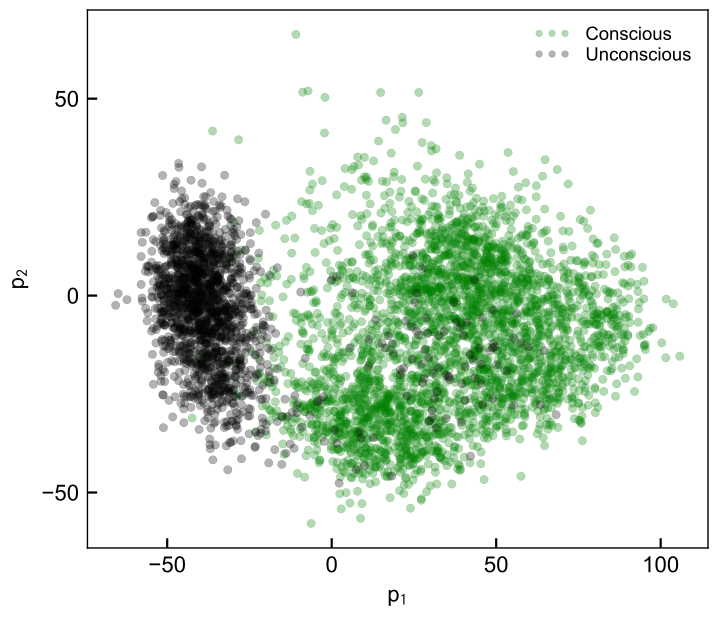}    
\caption{Example of $p_{1,t}$ (PC2 scores) vs. $p_{2,t}$ (PC3 scores) plotted for one subject, demonstrating the feasibility of using values of $p$ as targets for control that ensure unconsciousness.} 
\label{fig:cu}
\end{center}
\end{figure}

We eliminated PCs above PC3, which each contained $<2$\% of the observed variance in the spectrogram.
We chose to use two PCs ($p_1$ = PC2, $p_2$ = PC3) due to their clear concentration-dependence and good dynamic range, thus enabling control to be applied.
We did not use PC1 due to its unclear dependence on concentration.
Furthermore, these two PCs correspond with unconscious or conscious state, as shown by the separation between conscious and unconscious signal in Figure~\ref{fig:cu}, and thus enable a controller to use a set point relative to conscious state.

\subsection{State-space dynamics and inter-individual variation}
Having reduced the dimensionality along which the PSD evolves during anesthesia, we next sought to characterize pharmacodynamics (PD) by parameterizing a function $h_i(c)$ that generates $p_i(c)$.
Each $h_i(c)$ varies between patients (as seen in Figure~\ref{fig:pca}), and parameterization of these functions is important for performing control.
We sought to find a universal parameter set so that control may be implemented for any individual, and so $h_i(c) = h_i(c;a_i)$ where $a_i$ is a fixed parameter set for function $h_i$, common to all subjects.

\begin{figure}
\begin{center}
\includegraphics[width=8.4cm]{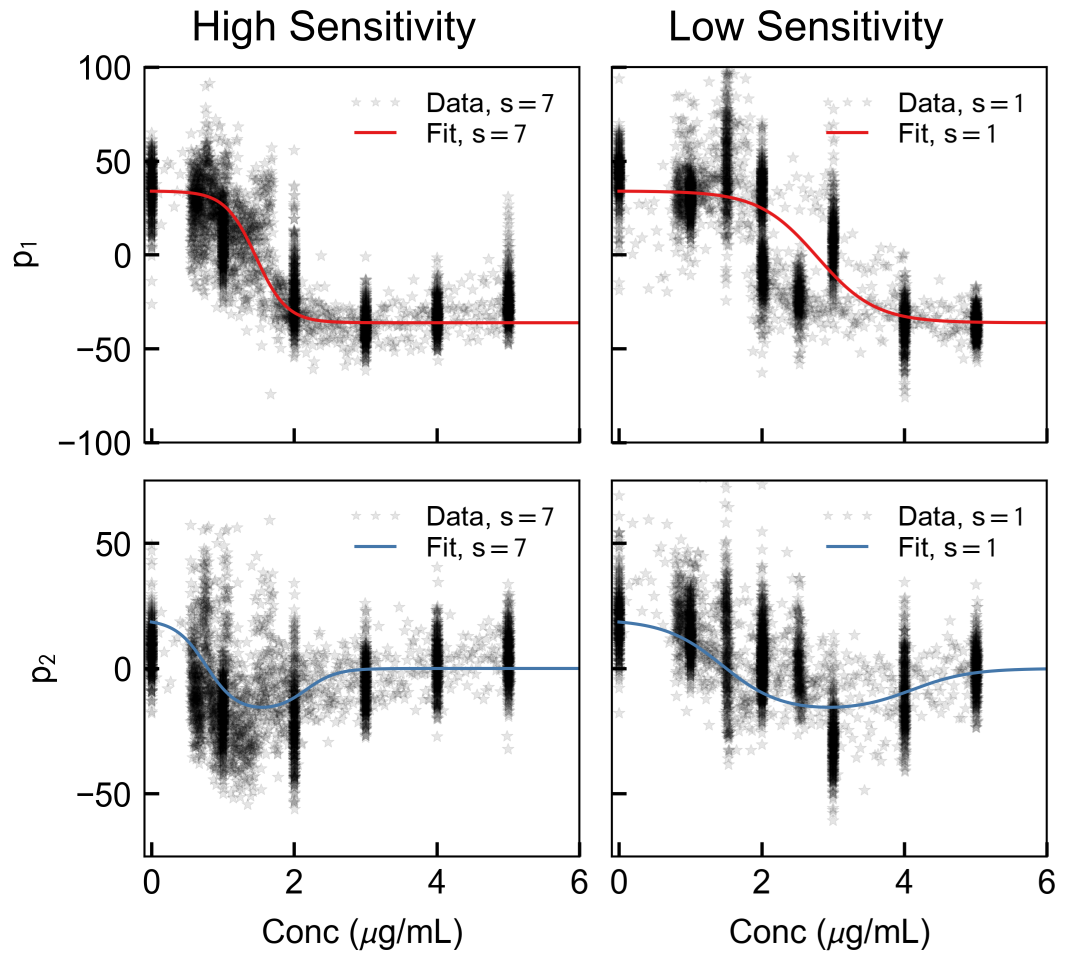}    
\caption{Example of two-dimensional control signal identified for two representative subjects, one with a high anesthetic sensitivity (left, $k_7=4.63$) and one with a low anesthetic sensitivity (right, $k_1=2.44$).} 
\label{fig:fit}
\end{center}
\end{figure}

Ideally the PD model we developed would apply without any subject-specific tuning.
Although the shapes of $h_i(c)$ are consistent across individuals, they appear to be scaled by a constant parameter (Figure~\ref{fig:fit}, top).
Because we have $S=10$ subjects, we can then state that for subject $s$, 
\begin{equation}
h_i(c) = h_{i,s}(c;a_i,k_s) = h_{i,s}(k_s c;a_i)
\end{equation}
where $k_s$ is a the subject's anesthetic sensitivity
and a high $k_s$ denotes high sensitivity.
A higher sensitivity indicates a larger response to a given anesthetic dose.
This is of additional use because, since $k_s$ is always a coefficient of $c$, 
\begin{equation}
\frac{dp}{dc} = \frac{d}{dc} h(k_s c) = k_s h'(k_s c) 
\end{equation}
where $h'(c) = \partial h/\partial c$.
Thus the equation we seek to control for subject $s$ becomes:
\begin{equation}
\frac{dp_s}{dt} = k_sh'(k_sc)\left(\frac{dc}{dt}\right).
\end{equation}
By parameterizing $h_1$ and $h_2$ and taking the derivative, we found an analytic form of $dp/dt$ which depends on the anesthetic sensitivity.

By inspection, we postulated that these functions may be captured with logistic equations, specifically:
\begin{equation}\label{eq:hdef} 
\begin{split}
h_{1,s}(k_s c ; a_1) =& \frac{a_{1,1}}{1+\exp(k_sc - a_{1,2})} - a_{1,3}\\
h_{2,s}(k_s c ; a_2) =&  \frac{a_{2,1}}{1+\exp(k_sc - a_{2,2})} - \frac{a_{2,3}}{1+\exp(k_sc - a_{2,4})}.
\end{split}
\end{equation}
Thus, parameter set $a_1\in\mathbb{R}_{\geq0}^3$, parameter set $a_2\in\mathbb{R}_{\geq0}^4$, and patient-specfic parameter set $k = {k_1,\cdots, k_s}\in\mathbb{R}_{\geq0}^S$.

We used a least-squares fitting to parameterize $h$, that is:
\begin{equation} \label{min}
\begin{split}
A = \arg\min_{a_1, a_2, k}&\sum_i\sum_s\sum_t ||(h_{i,s}(k_s c_t ; a_i) - p_{i,t,s})||_2\\
&\text{subject to:} \\
&\text{Eqn. \eqref{eq:hdef}}
\end{split}
\end{equation}
This optimization resulted in parameters 
\begin{equation}
\begin{split}
a_1=&[70.1,\,6.8,\,36.2],\\
a_2=&[37.0,\,3.6,\,17.4,\,10.0],\\
k = &[2.44,\,6.78,\,3.22,\,2.88,\,3.55,\\
    &\,\,\,5.79,\,4.63,\,2.97,\,4.15,\,2.61],
\end{split}
\end{equation} as shown in Figure~\ref{fig:fit}.
Finally, taking $dh/dc$ yields our PD control model in differential equation form:
\begin{equation}
\begin{split}
\label{eq:diffeq}
&\frac{dp}{dc} = \frac{d}{dc} 
\begin{bmatrix}
h_{1,s}\\
h_{2,s}
\end{bmatrix}
\\
&=k_s
\begin{bmatrix}
\frac{-a_{1,1}\exp(k_sc - a_{1,2})}{(1+\exp(k_sc- a_{1,2}))^2}\\
\frac{-a_{2,1}\exp(k_sc - a_{2,2})}{(1+\exp(k_sc - a_{2,2}))^2} + \frac{a_{2,3}\exp(k_sc-a_{2,4})}{(1+\exp(k_sc-a_{2,4}))^2}
\end{bmatrix}.
\end{split}
\end{equation}
We note that $k_s$ is still a subject-specific parameter which we do not know \textit{a priori}.
Luckily, we may estimate and update this parameter in the same way an anesthesiologist may adjust dosing to determine an individual's anesthetic sensitivity.

\subsection{Incorporating pharmacokinetics}
There are numerous PK models used under varying circumstances in the operating room \citep{Schnider1998, Schnider1999, Levitt2005, Barakat2007, Absalom2009}.
These are most generally two- or three-compartment models, but may be increasingly complicated and multicompartmental.
Model parameters are also typically a function of patient body mass and age \citep{Schnider1998, Schnider1999}.
The model from \cite{Schnider1999} was used in \cite{Purdon2013} to provide the $c$ values we have used in this study, directly supporting the utility of this approach in conjunction with the methods we have developed.
For our control implementation here, we used the same four-compartment PK model from \cite{Schnider1998, Schnider1999} and used in \cite{Chakravarty2017}:
\begin{equation}\label{eq:pk}
\begin{split}
\frac{d}{dt}
\begin{bmatrix}
c_1\\
c_2\\
c_3\\
c
\end{bmatrix}
&=\\
&\begin{bmatrix}
-k_{12}-k_{10}-k_{13}-k_{1e} & k_{21} & k_{31} & k_{e1}\\
k_{12} & -k_{21} & 0 & 0\\
k_{13} & 0 & -k_{31} & 0\\
k_{1e} & 0 & 0 & -k_{e1}
\end{bmatrix}
\begin{bmatrix}
c_1\\
c_2\\
c_3\\
c
\end{bmatrix}\\
&+
\begin{bmatrix}
1\\
0\\
0\\
0
\end{bmatrix}
u(t)
\end{split}
\end{equation}
where $u$ is the anesthetic infusion concentration, $c_{1,2,3}$ are non-effect compartment concentrations, and $c$ is the effect site concentration all in $\mu g/mL$.
We used parameters corresponding to a 24 year old female patient with a mass of 65kg and a height of 163cm.
We note that this would be replaced with a model parameterized by patient characteristics in a real-world implementation.

\section{Implementation of Control}

\subsection{Formulating the control problem}
The model developed in the prior sections may now be used to formulate an \textit{in silico} control problem to test the feasibility of this approach.
The nonlinear nature of the control equations lends itself to a nonlinear model predictive control (NMPC) approach. 
NMPC has been used in biological applications previously, with good successes \citep{Abel2019, Dassau2017}.
In this case, we set desired $p = [p_1,\,p_2]$ and control the system to remain stably at those values.
In the future, the desired $p$ values may be selected by training classifiers using the conscious state of the subject (e.g., sedation may correspond to one $p^{sed}$, whereas general anesthesia might correspond to another $p^{ga}$).

\begin{figure}
\begin{center}
\includegraphics[width=8.4cm]{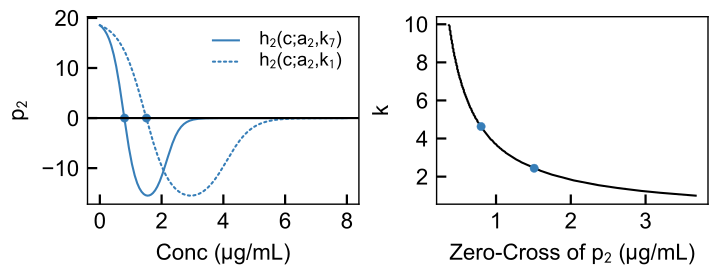}    
\caption{Finding $k_s$ is enabled by the dependence of the zero-cross of $p_2$ on $k_s$. Left, $h_2(c;a_2,k_s)$ is plotted for two values of $k_s$. The initial zero-cross of these depends upon $k_s$. Given the parameterization of $h$ we have developed, the relationship between $k_s$ and the PK effect-site concentration at the predicted zero-cross of $p_2$ is shown here. Thus, by updating the controller parameterization when $p_2$ is observed to cross zero, we may attain individualized control.} 
\label{fig:nadir}
\end{center}
\end{figure}

We still retain the problem of the unspecified $k_s$.
In practice, a standard induction bolus is given to each patient, and the dosing is then modified depending upon how the patient responds.
In the same manner, we may initialize the controller with a set $k_s$ value, and update this value as we observe how the system responds.
One possible marker for identifying $k_s$ is the zero-cross of $p_2$ (PC3), as it varies systematically with $k$ and is observed prior to deep unconsciousness.
Figure~\ref{fig:nadir} shows the relationship between the observed zero-cross of PC3 and $k$, given the parameterization we have previously identified.

The state of the patient updates every 2 s, given the multitaper parameters we have selected.
Anesthetic infusion pumps update more infrequently, so we parameterized $u$ as piecewise-constant with 10 s steps with control input bounded on $[0, u_{max}]$.
The anesthetic takes several minutes to deliver its effect, and so we used control and prediciton horizons of 300s ($M = 30$ steps of 10 s) with the predicted value $p^{pred}$ computed at the end of each of the steps.
We initialized the model with an initial sensitivity $k_{init} = 2$ and update this sensitivity once $p_2$ crosses zero.
The predicted states starting at time $t_0$ are given by integrating Eqn. \eqref{eq:diffeq} forward 10 s at a time.
\begin{equation}
\label{eq:pred}
\begin{split}
p^{pred}_{m+1}(u_m; a, k_{e}) &=\\
 p^{pred}_m &+ \int_0^{10} \left(\frac{dh}{dc}(c;a,k_e)\right)\left(\frac{dc}{dt}(c,u_m)\right) dt
 \end{split}
\end{equation}
where $p^{pred}_m$ is the value at the end of the previous step and $p^{pred}_{m=0}$ is the current observed state of the system.
We denote the current estimate of $k_s$ as $k_{e}$, which is initialized at a value of $k_{init}=2$ and updated following the zero-cross of $p_2$.

Thus, we formulated the NMPC finite-horizon optimal control problem for finding the optimal control $u_{\rm MPC}^\star$ over the predictive horizon as follows:
\begin{eqnarray}
    \nonumber  u^\star_{\rm MPC} &= \arg\min\limits_{u} \;\sum_{m=1}^{M} w_p||p^{pred}_m-p^{ga}|| +w_u||u_m||\\
        \label{eq:mpcoptim} &\text{subject to:}\\
        \nonumber &\text{Eqn. \eqref{eq:diffeq}}\\
        \nonumber &\text{Eqn. \eqref{eq:pk}}\\
        \nonumber &\text{Eqn. \eqref{eq:pred}}\\
        \nonumber  &0 \le u_{m}\le u_{\max},
\end{eqnarray}
where $w_p=1$, $w_u=0.0001$ to scale the minimization and $u_{max}=10$mg/kg/min.
We chose $p^{ga} = h(10; a, k_{init})$ because this corresponds to where the system is definitely unconscious (projected value for a high effect site concentration).
We are prevented from overdosing the patient (the trivial solution to ensuring unconsciousness) by the tunable control input penalty and the restriction on pump concentration rate.

We applied this NMPC controller to a model of the system given by:
\begin{equation}
\begin{split}
\frac{dp}{dt} =& \left(\frac{dh}{dc}(c;a,k_s)\right)\left(\frac{dc}{dt}(c,u)\right)\\
p(t=0) =& h(0;a,k_s)
\end{split}
\end{equation}
and note that we have isolated system-model mismatch to the $k$ parameter and ignored measurement noise.
Further testing would be needed to determine robustness of this approach to other parameter errors and observer design.

\subsection{Two \emph{in silico} examples}
First, we simulated an example where the individual is more sensitive to the anesthetic than the initial sensitivity $k_s=2k_{init}$.
The risk in such a case is unintentional overdosing of a patient.
This can be avoided by simple bounds on control input $u$, which may be relaxed or tightened after $k_s$ is found.
Next, we simulated an example where the individual is less sensitive to the anesthetic than the initial sensitivity $k_s=0.75k_{init}$.
The risk in such a case is underdosing of anesthesia and the patient retaining consciousness.

\begin{figure*}
\begin{center}
\includegraphics[width=16cm]{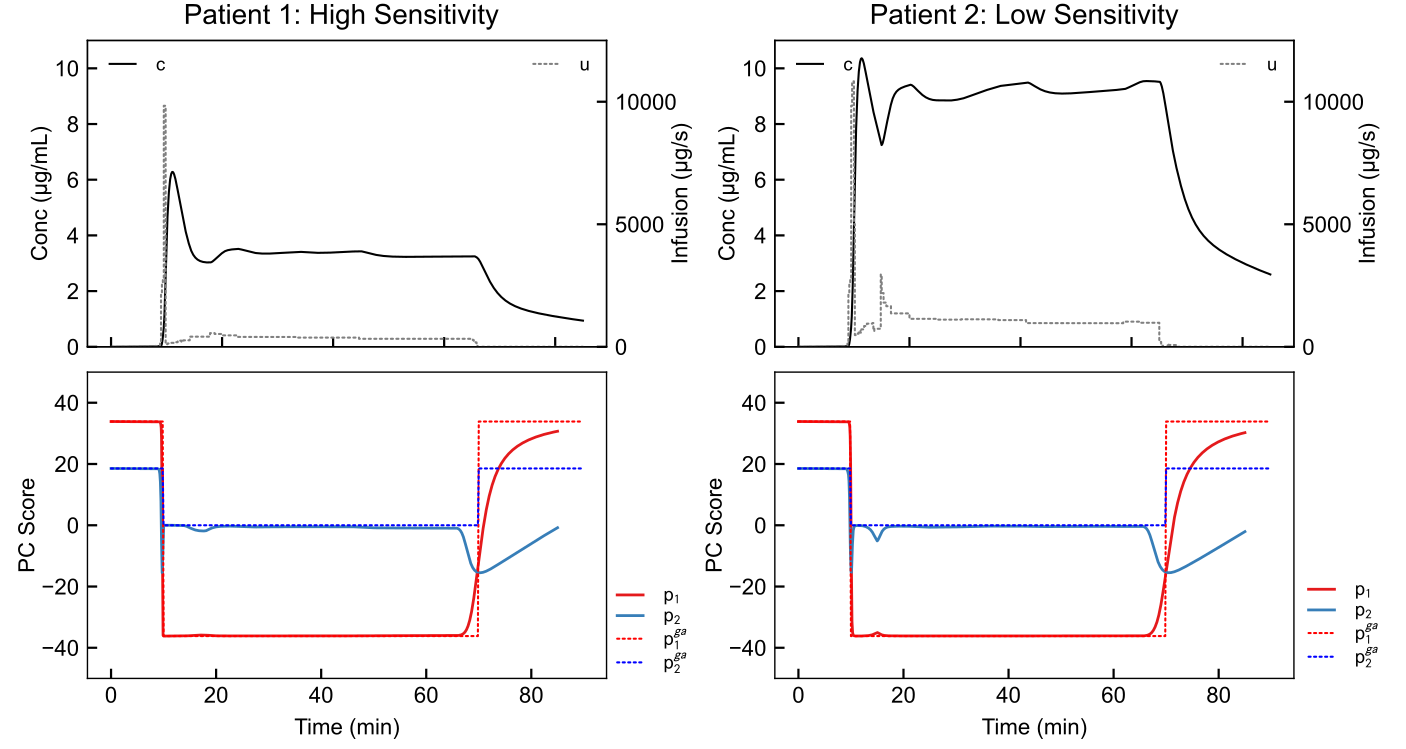}    
\caption{NMPC results for high anesthetic sensitivity $k_s=2k_{init}$ (left), and low anesthetic sensitivity, $k_s=0.75k_{init}$ (right). In this simulation, the set point moves from conscious (first ten minutes) to unconscious (next 60 minutes) to conscious (final 20 minutes). The controller in each case responds by inducing unconsciousness starting at the 10 minute mark. Furthermore, it correctly maintains unconsciousness using a lower anesthetic dose in the high-sensitivity patient.} 
\label{fig:control}
\end{center}
\end{figure*}

The results of these simulations are shown in Figure~\ref{fig:control}.
We found that this controller design performs well in generating a signal corresponding to unconsciousness in our model without excess or insufficient delivery of the anesthetic, i.e., it is sensitive to patient characteristics.
Both $k_s$ values (1.5, 4.0) are extremes near the range observed in the clinical study of healthy volunteers.
A large difference in effect-site concentration between these two cases (despite all patient characteristics remaining identical) resulting in the same PD underlines the need for PD-based control, rather than simple control of plasma or effect-site concentration.
Further testing of response to signal noise, system-model mismatch, and disturbance rejection is necessary, however, these simple examples provide a proof-of-concept.

\section{Discussion}
Several additional steps should be taken to validate the control signal we have selected.
First, the signal should be analyzed in clinical cases to determine if it follows the same dynamics in the presence of other drugs that may affect EEG.
Additionally, there are deeper states of anesthesia, such as burst-suppression, which should be avoided if they are not clinically desired \citep{Brown2010}.
Currently, the controller avoids deeper-than-needed anesthesia by minimizing the control input needed to attain the depth of anesthesia setpoint, however, if the control signal continues to evolve as anesthesia deepens, a control signal corresponding to burst-suppression may be avoided explicitly.
Characterization of these states may be achieved using data recorded during clinical administration of anesthesia in the operating room \citep{Purdon2015}.
Running the controller retroactively on EEG recorded during clinical cases and comparing controller suggestions with anesthesiologist action is a reasonable next step in testing the control system we have presented.
We note that observer design will also be important in applying control in this fashion. 

Despite these barriers, there are two main benefits to using NMPC to control this system in comparison to LQR or PID controllers.
First, NMPC does not attempt to use a linear approximation of model dynamics, and therefore requires less simplification of the complex underlying neural system.
Second, safety mechanisms may be readily implemented in NMPC.
We anticipate that a clinical NMPC system would involve safety features such as anesthesia-on-board constraints to prevent overdosing (as in \cite{Ellingsen2009}) and modifying controller responsiveness via confidence indexes (as in \cite{Pinsker2018, Laguna2017}).

Significant barriers remain to realizing closed-loop control of general anesthesia in clinical settings \citep{Absalom2011a}.
By developing physiologically interpretable PD models of anesthesia, we seek to bridge the gap in understanding between the anesthesiologist and the controls engineer.

\bibliographystyle{myIEEEtran}
\bibliography{condensed_library}

\end{document}